\documentclass[11pt,twoside]{article}

\usepackage{asp2006}
\usepackage{epsf}
\usepackage{psfig}
\usepackage{lscape}
\usepackage{graphicx}

\begin{document}
\title{Towards Asteroseismology of the Multiperiodic Pulsating
Subdwarf B Star PG1605+072} 
\author{A. Tillich,$^1$ U. Heber,$^1$ S. J. O'Toole,$^2$ 
R. \O{}stensen,$^3$ and S. Schuh$^4$}
\affil{$^1$Dr.Remeis-Sternwarte Bamberg, Universit\"at
Erlangen-N\"urnberg, Sternwartstr.7, D-96049 Bamberg,
Germany}
\affil{$^2$Anglo-Australian Observatory, P.O. Box 296 Epping,
NSW 1710, Australia}
\affil{$^3$Instituut voor Sterrenkunde,
Celestijnenlaan 200D, 3001 Leuven, Belgium}
\affil{$^4$Institut f\"ur
Astrophysik, Universit\"at G\"ottingen, Friedrich--Hund--Platz~1, 37077
G\"ottingen, Germany}

\begin{abstract}
We present our attempt to characterise the frequency spectrum of the
bright V 361 Hya star PG1605$+$072 from atmospheric parameter and
radial velocity variations through a spectroscopic approach. Therefore
we used time resolved spectroscopy ($\approx$9000 spectra) to detect
line profile variations from which variations of the effective
temperature and gravity, together with their phases, are extracted by
means of a quantitative spectral analysis.  The theoretical modelling
of adequate pulsation modes with the \textit{BRUCE} and \textit{KYLE}
codes allows to derive constraints on the mode's degree and order from
the observed frequencies.
\end{abstract}

\section{Introduction}
Asteroseismology provides a promising avenue to determine stellar
masses.\linebreak Among the sdBs two classes of multi-mode pulsators
are known, the V361~Hya and V1093 Her stars.  The V361 Hya ones are of
short period (2--8\,min), while the V1093 Her stars have longer
periods (45 to 120\,min).  The short period oscillations can be
explained as acoustic modes ($p$-modes) of low degree $l$ and low
radial order $n$ excited by an opacity bump due to a local enhancement
of iron-group elements (Charpinet et al.\ 1997).  PG1605$+$072 (also
known as V338 Ser) was discovered to be a V361\,Hya star by Koen et
al.~(1998) and was found to have the longest periods (up to 9\,min) of
this class of stars and a very large photometric amplitude (up to
$\sim$60\,mmag).  They detected a very strong main mode together with
about 20 other modes.  The first multi-site campaign confirmed these
results and increased the number of observed frequencies to 50
(Kilkenny et al.\ 1999).  Using high resolution Keck spectra, the
stellar parameters and metal abundances were derived by Heber et
al.~(1999): $T_{\rm{eff}}=32\,300$\,K, $\log{(g
[\mathrm{g}\,\mathrm{cm}^{-2}])}=5.25$\,dex, and
$\log{(n_\mathrm{He}/n_\mathrm{H})}=-2.53$ whereas $n$ is the number
density (cgs-units are used henceforth). The small $\log{g}$ implies that
the star is quite evolved and has already left the Extreme Horizontal
Branch (EHB). Radial velocity variations due to pulsation were
detected for the first time by O'Toole et al.~(2000). We organised a
coordinated multi-site spectroscopic campaign to observe PG1605$+$072
with medium resolution spectrographs on 2\,m and 4\,m telescopes.

\section{The Analysis of the MSST Data}
During the Multi-Site Spectroscopic Telescope (MSST) campaign, we
obtained 151 hours of time-resolved spectroscopy on PG1605+072 at four
observatories in May/June 2002 (O'Toole et al.\ 2005, henceforth
Paper~I).  The spectra obtained at Steward Observatory outnumber the
data from the other observatories and are of much better S/N than the
rest of the data sets. Hence we focus mainly on the results from the
Steward data and refer to Tillich et al.\ (2007, henceforth Paper II)
for a complete analysis.  The entire data set has already been
analysed for radial velocity variations (Paper~I) and 20 modes have
been detected with radial velocity amplitudes between 0.8 and
15.4\,$\mathrm{km}\,\mathrm{s}^{-1}$.
\begin{figure}[ht!]
\centering
\includegraphics [angle=270,scale=.575,bb=302 107 554 740] {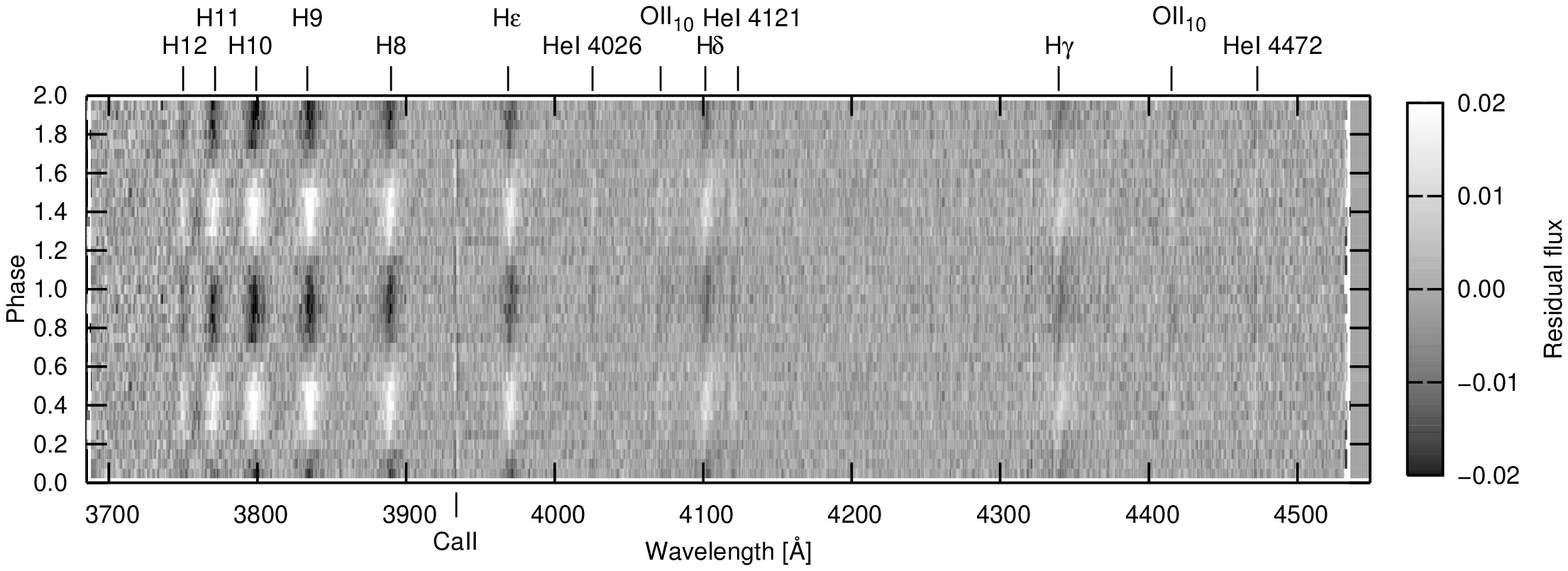}
\includegraphics [angle=270,scale=.525]{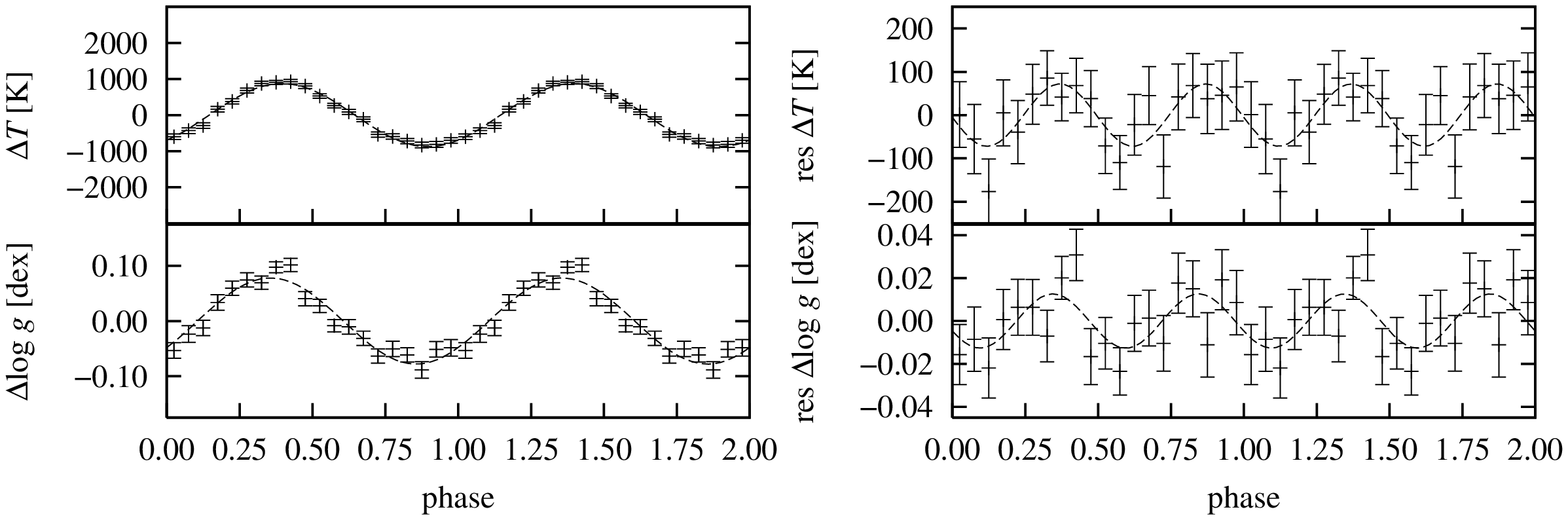}
\caption{\textit{Top:} Line profile variations of the phase binned
Steward data for the strongest mode $f1$ (period P=481.74\,s). The Ca\,II
line is of interstellar nature. Its variation is due to the stellar
radial velocity correction. Besides the strong Balmer, He\,{\sc i}\ and
O\,{\sc ii}\ lines are found to vary (from Tillich et al. 2007).
\textit{Bottom:} Variations of the temperature, log $g$ (both left),
temperature and log $g$ residuals (both right) with best fit sine
curve and statistical error bars.}\label{fig_f1}
\end{figure}
As individual spectra are too noisy for a quantitative spectral
analysis in terms of detecting very small variations, they were
combined in an appropriate way.  To be able to detect tiny variations
for any pre-chosen pulsation mode, we determined the phase of each
individual spectrum according to the selected pulsation period and
co-added them weighted by S/N.  To this end, a complete pulsation
cycle was divided into twenty phase bins.  In Fig.~\ref{fig_f1}
(\textit{top}), the phase dependent changes in the line profiles of
the Steward data with respect to the mean spectrum are shown.  It is
evident that all the observed H, He~{\sc i}\ and O {\sc ii}\ lines
vary in the same way with phase.  Effective temperatures
($T_{\rm{eff}}$), surface gravities ($\log{g}$), and helium abundances
($\log{n_\mathrm{He}/n_\mathrm{H}}$) were determined for every bin by
fitting synthetic model spectra to all hydrogen and helium lines
simultaneously, using a procedure developed by Napiwotzki et
al.~(1999).  A grid of metal line-blanketed LTE model atmospheres with
solar metal abundance was used.

\begin{figure}[ht!]
\centering
\includegraphics [scale=.3]{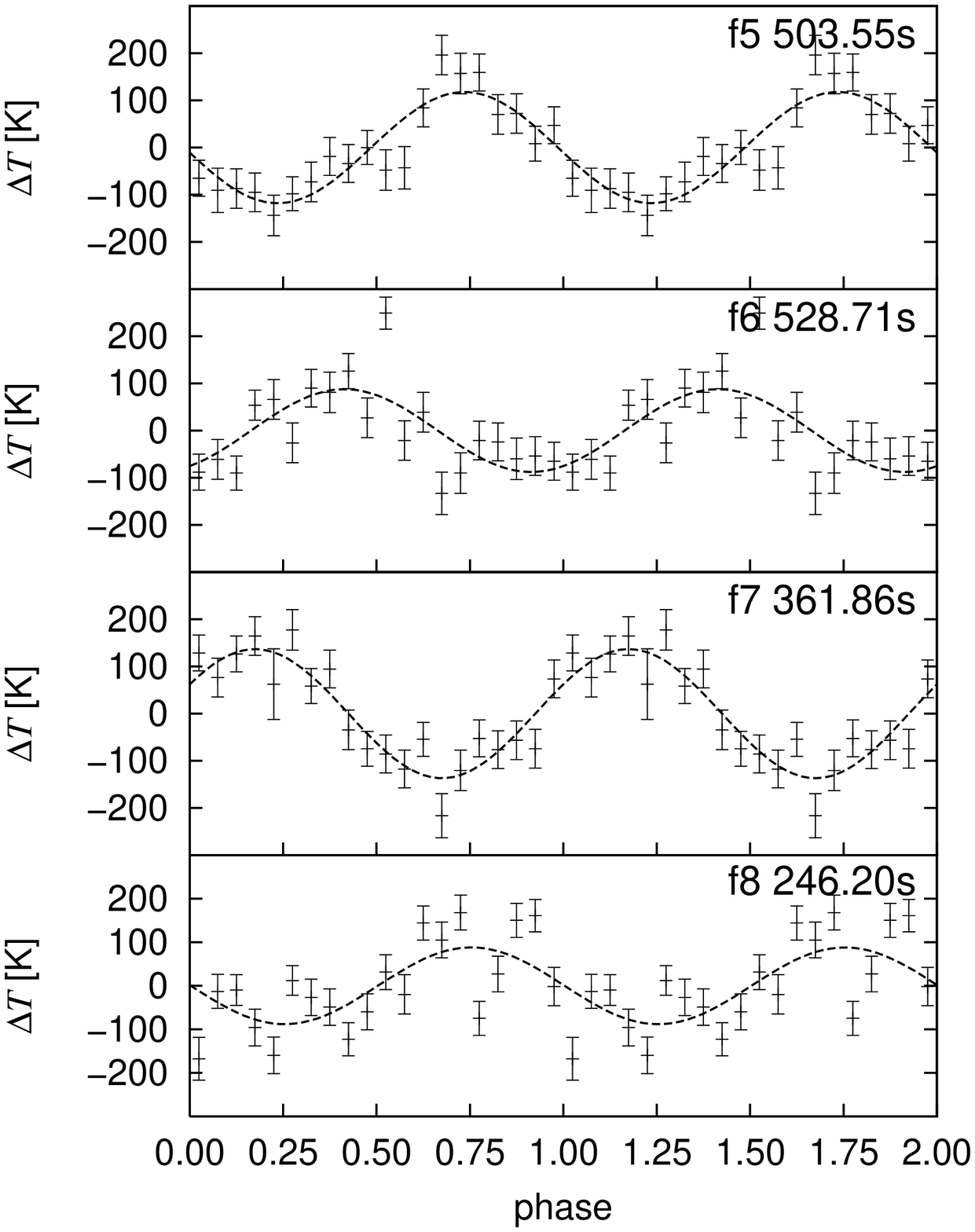}
\includegraphics [scale=.3]{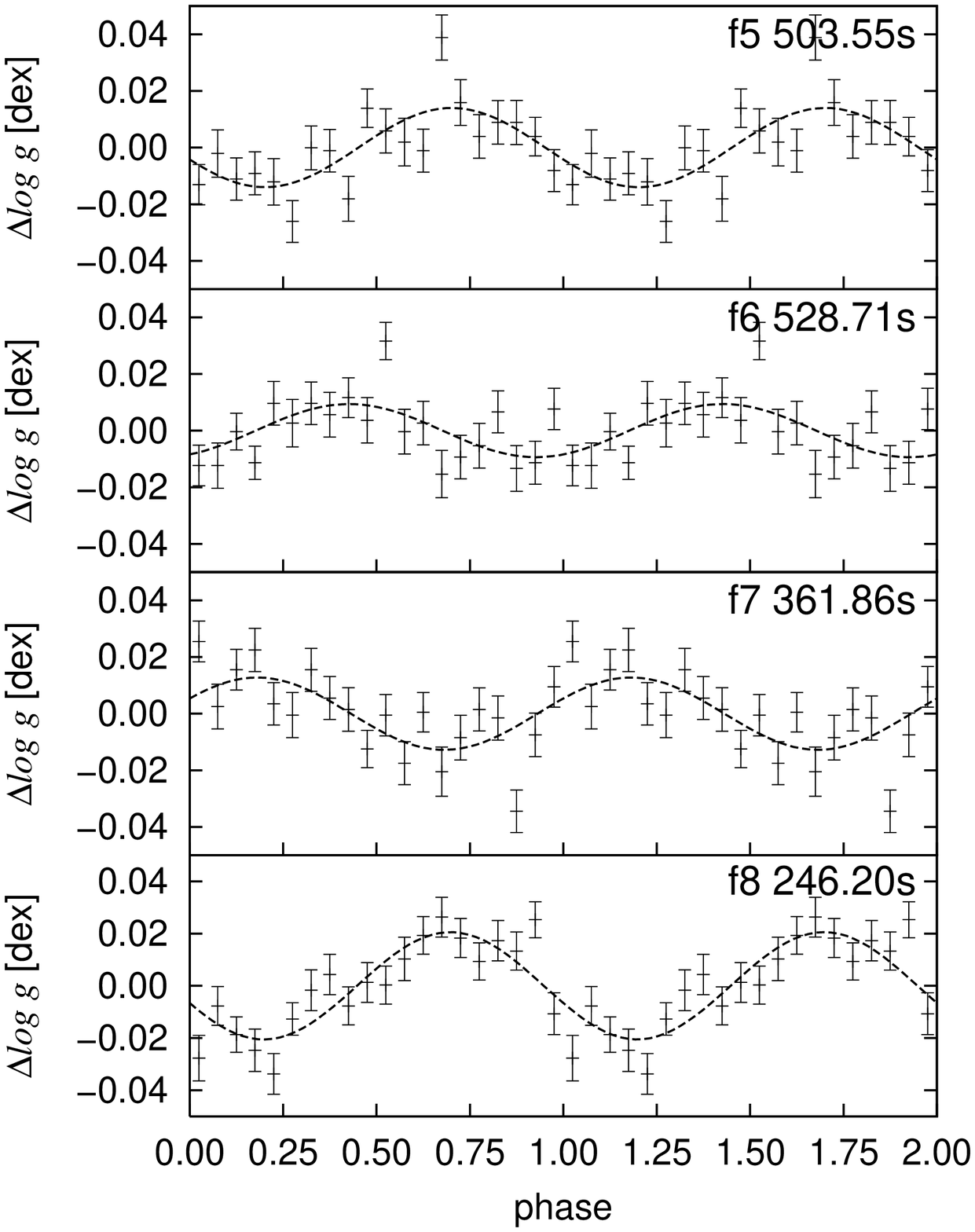}
\caption{Variations of the temperature ({\it left}) and the log $g$ 
({\it right})
for the modes $f5$, $f6$, $f7$ and $f8$ after the cleaning procedure with sine
fits and statistical error bars derived for the Steward data (from
Paper II).}\label{fig_weakmodes}
\end{figure}

We start our analysis with the dominant mode in radial velocity, which
is expected to also show the largest variation in temperature and
gravity (see Fig.~\ref{fig_f1}, \textit{bottom}).  The variations of
$T_{\rm{eff}}$ and $\log{g}$ are sinusoidal. Therefore, the pulsation
semi-amplitude is determined by using a $\chi^2$ sine fitting
procedure ($\Delta T_{\rm{eff}}=873.7\pm16.5$ and
$\Delta\log{g}=0.078\pm0.003$). As expected the helium abundance
remains constant. The first harmonic is clearly recovered in the
temperature residuals (see Fig.~\ref{fig_f1} right, \textit{bottom}).
We then proceeded to search for and analyse weaker frequencies for
which
\begin{figure}[ht!]
\centering
\includegraphics [scale=.30]{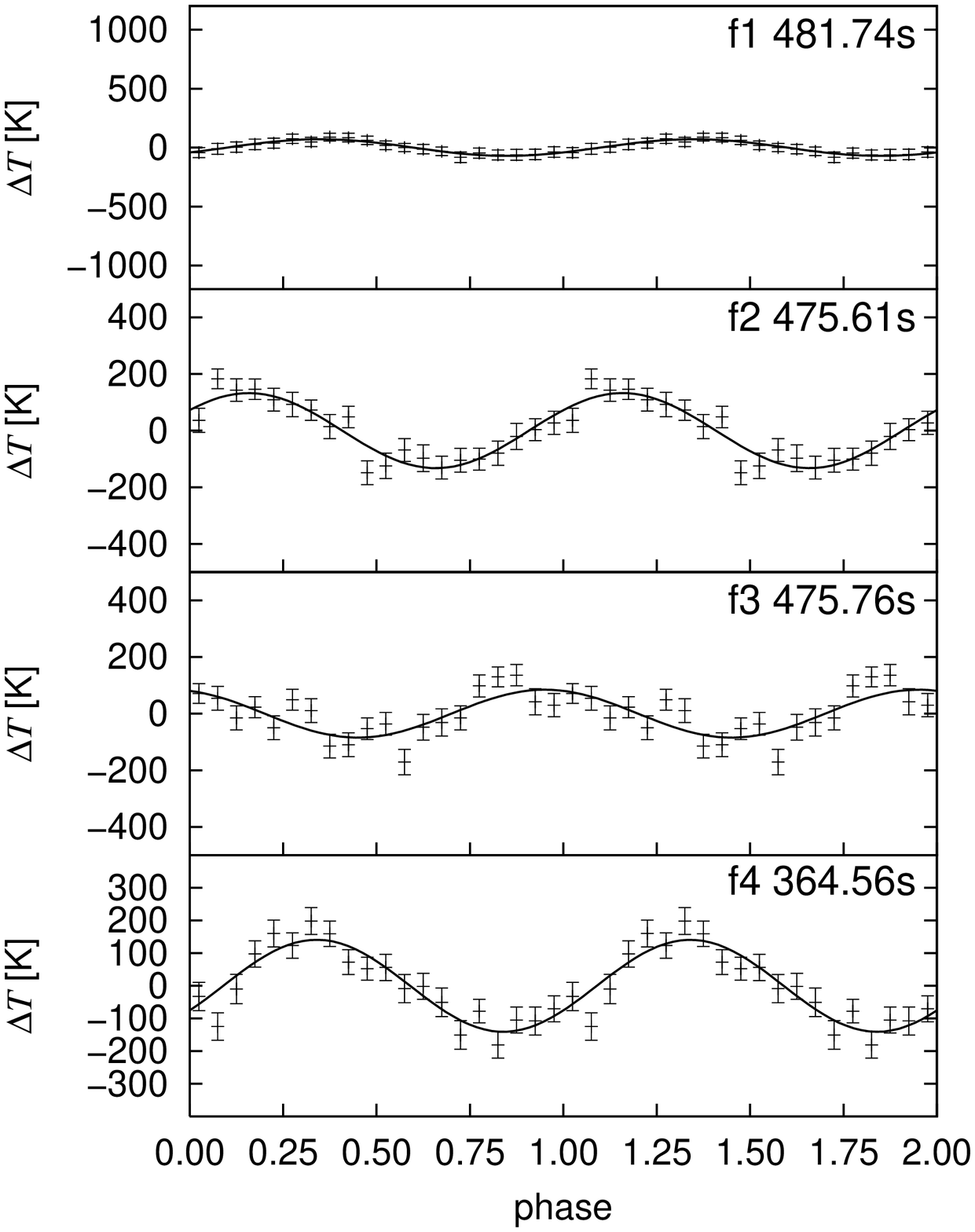}
\includegraphics [scale=.30]{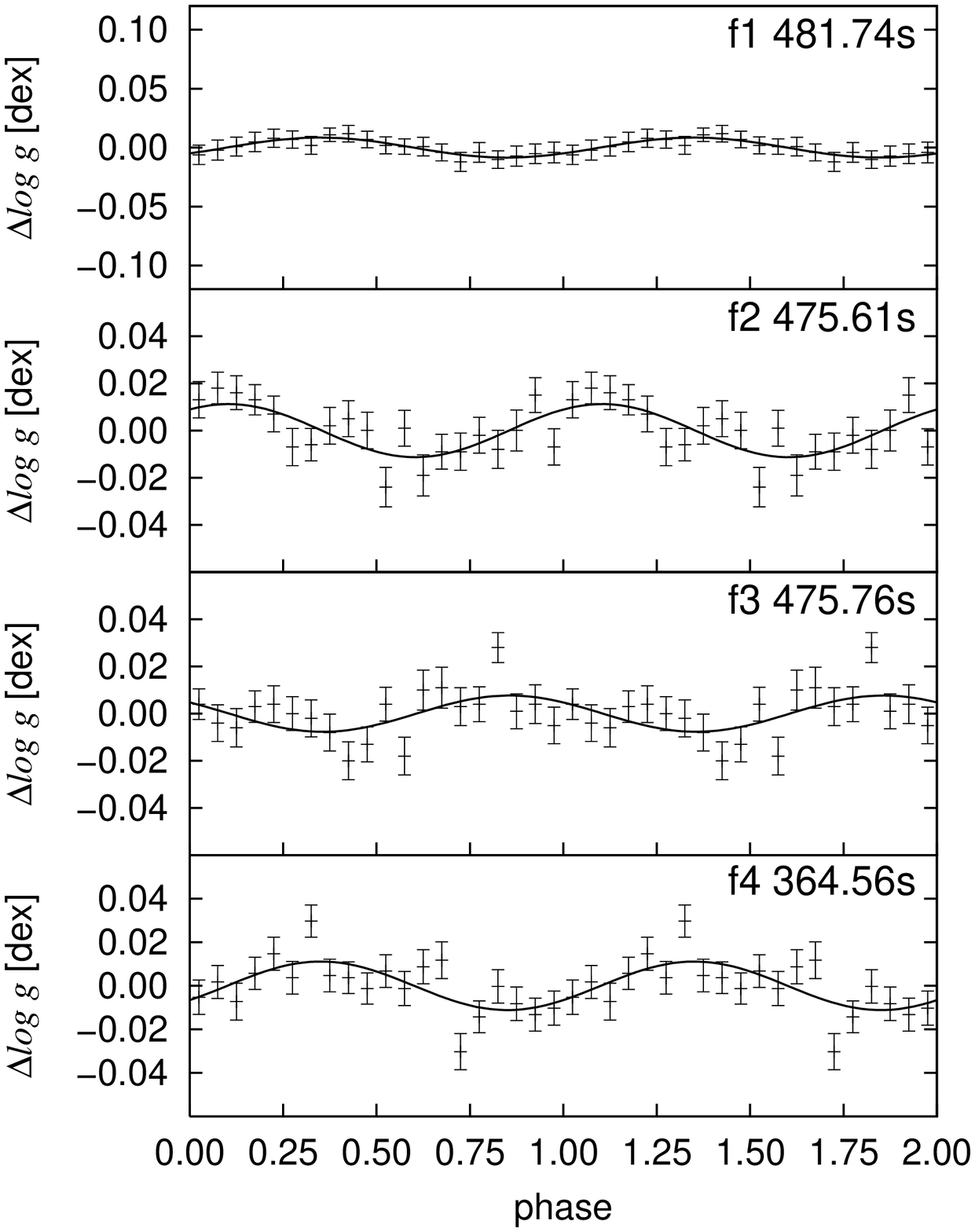}
\caption{Variations of the temperature
({\it left}) and the log $g$ ({\it right}) for the four strongest frequencies
after the cleaning procedure with the statistical error bars. The best
fit sine curves are the full drawn lines. Note that the dominant mode
is not perfectly eliminated. However the residuals are very
small.}\label{fig_f124}
\end{figure}
we developed a cleaning procedure.  First we calculated synthetic
spectra for every phase bin of the dominant pulsation frequency
$f1$. Then they were summed up to create the mean synthetic spectrum for
$f1$, which was subtracted from each phase bin spectrum to form the
cleaning function.  This correction was applied to all of the $\sim
9000$ individual spectra by subtracting the cleaning function for the
corresponding phase of the dominant mode. All individual cleaned
spectra were then summed into the appropriate phase bin for the
corresponding periods of lower amplitude frequencies (from $f2$ onwards)
and were analysed for the atmospheric parameter variations as
described above. Results are summarized in Table 1 and plotted in
Figs.~\ref{fig_f124} and~\ref{fig_weakmodes}.  Unlike the isolated
frequencies $f4$, $f2$ and $f3$ are too close together and therefore
unresolved in our analysis.  For the frequency $f9$, the errors of our
data points ($\delta
T_{\rm{eff}}=40\,\rm{K},\delta\log{g}=0.008\,\rm{dex}$) are of the
same order of magnitude as the amplitude ($\Delta
T_{\rm{eff}}=35\,\rm{K},\Delta\log{g}=0.003\,\rm{dex}$) and the
detection of variations must be regarded as marginal.  Nevertheless,
this demonstrates that it is possible to reveal variations of
atmospheric parameters of modes with radial velocity variations as
small as $2\,\mathrm{km}\,\mathrm{s}^{-1}$.  Frequencies $f5$ and $f6$ are
isolated and therefore well resolved.  By contrast frequencies $f7$, $f9$
and $f11$ are close together and therefore unresolved. While amplitude
variations of atmospheric parameters are found for $f7$, the variations
for $f9$ and $f11$ are below the detection limit.  $f8$ and $f10$ are
combination frequencies involving $f1$ ($f1$+$f5$ and $2\times f1$,
respectively). For $f8$, the variation of $T_{\mathrm{eff}}$ is detected
only marginally, whereas the gravity variation is pronounced (see
Fig.~\ref{fig_weakmodes} and Table~1).  In order to characterise
pulsation modes, it is also important to investigate the phase lags
between temperature, gravity and radial velocity variations.  These
phase lags can be used to determine the deviations from an adiabatic
change of condition.  The radial velocities of individual spectra were
taken from Paper I.

The amplitudes of their variations were determined in the same way as
the temperature and gravity variations using the phase binning and
cleaning technique described in Sect.~2. Phase lags between the
variation of radial velocity and the atmospheric parameters can be
derived directly by comparing the phases of their maxima.  In order to
determine the phase lags of the weaker modes, we used the temperature
curves, and measured just the phase lags for two sine functions.  As
the formal fitting errors are unrealistically small, we use half the
bin size ($\pm 0.025$) as error estimate.  It is worth noting that for
all recovered modes (except $f3$), the radial velocity variation reaches
its maximum before the temperature variation. Furthermore, all the
values of the phase lags lie around $0.25$($\widehat{=}\frac{\pi}{2}$)
with small but significant deviations. This is the value we would
expect for a completely adiabatic $p$-mode pulsation.
\begin{table}[ht!]
\caption{Semi-amplitudes of the temperature and gravity
variations after cleaning for the four strongest modes derived from
Steward data. Also the statistical error from the $\chi^2$ sine
fitting is listed. Periods, frequencies and radial velocity amplitudes
are taken from Paper I and therefore calculated for the whole MSST
data run.}
\centering\smallskip
\begin{tabular}{cccccc}
\tableline
\noalign{\smallskip}
Name & Period & $f$ & $v$ &  $\Delta T_{\rm{eff}}$ & 
$\Delta\log{g}$ \\
& [s] & [$\mu$Hz] & $[\mathrm{km}\,\mathrm{s}^{-1}]$ & [K] & [dex] \\
&     &           &            & cleaning & cleaning \\ 
\noalign{\smallskip}
\tableline
\noalign{\smallskip}
$f1$ & 481.74 & 2075.80 & 15.429 &  70.5$\pm$4.7   & 0.008$\pm$0.001\\ 
$f2$ & 475.61 & 2102.55 & 5.372  &  132.6$\pm$12.6 & 0.011$\pm$0.002\\ 
$f3$ & 475.76 & 2101.91 & 2.971  &  84.5$\pm$17.5  & 0.008$\pm$0.002\\ 
$f4$ & 364.56 & 2743.01 & 2.497  &  140.9$\pm$14.6 & 0.011$\pm$0.002\\ 
$f5$ & 503.55 & 1985.89 & 2.474  &  117.9$\pm$10.3 & 0.014$\pm$0.002 \\ 
$f6$ & 528.71 & 1891.41 & 2.322  &  87.7$\pm$15.0  & 0.009$\pm$0.002\\ 
$f7$ & 361.86 & 2763.50 & 2.121  &  136.8$\pm$10.5 & 0.013$\pm$0.003\\ 
$f8$ & 246.20 & 4061.70 & 1.777  &  88.1$\pm$18.3  & 0.021$\pm$0.002\\ 
\noalign{\smallskip}
\tableline
\end{tabular}
\label{tab_pre}
\end{table}
But as a real star is a non-adiabatic system due to its radiation of
light, such deviations are to be expected.  For the dominant mode $f1$,
the phase lag between $T_{\rm{eff}}$ and $\log{g}$ is 0.3 (see Paper
II) indicating that the temperature is highest shortly after the
radius is smallest.  After that, we consider the phase lags between
the radial velocity and the gravity, which are supposed to have a
strict relationship, as they are both produced by the movement of the
stellar surface (i.e.\ the $g$ variation should be in line with the
derivative of the radial velocity curve). Hence, we expect a phase lag
of 0.25 between the RV curve and the $\log{g}$ curve.  We apply the
same fitting procedure as before. For the Steward data, most values
seem to be in agreement with the theory within
$\Delta\varphi=0.02$. This is consistent with our error estimate,
except for frequencies $f3$, $f7$ and $f8$. The difficulties with $f3$ have
already been discussed (see above) and the same may hold for $f7$ as it
is also unresolved from $f9$, whereas frequency $f8$ is close to the
detection limit (see above).

\section{The Modelling of Pulsation Modes}
For the modelling we used a routine, authored and introduced by
Townsend (1997). The adiabatic code \textit{BRUCE} has been developed
to model non-radial pulsations for early-type main sequence
stars. With a radiative shell and a convective core these stars have a
similar structure as subdwarf B stars. Therefore the routines should
also be able to describe the pulsations in these smaller sdB stars.
In order to model the surface variations by a certain pulsation mode,
the stellar parameters have to be known accurately, e.g.~the polar
values for radius, temperature, gravity and inclination as well as the
equatorial rotation velocity. From these assumptions an equilibrium
surface grid of the star, consisting of 51030 points, is calculated,
taking into account any effects of the rotation on both surface
geometry and surface temperature distribution. Thereafter this grid is
perturbed by one (or several) well defined pulsation mode(s) to reveal
the temporal propagation. The perturbation must be characterized
through degree and order of a spherical harmonic and a perturbation
velocity amplitude (for more details see Townsend 1997).  The
inclination angle plays a very important role for cancellation effects
except for radial modes.  \textit{BRUCE} provides us with all the
parameters neccessary for a spectral synthesis such as temperature,
gravity and the projected values for velocity, surface area and
surface normal for every point grid point on the stellar surface.  The
\textit{KYLE} routine is based on a code authored by Falter (2001)
and used to perform the spectral synthesis for every time step.  For
this purpose it interpolates in the same grid of LTE synthetic model
spectra used for the spectral analysis (see Sect.~2) to obtain a
spectrum for every point on the stellar surface. In order to shorten
the calculation time, our model atmosphere grid was restricted to nine
points, i.e. $30\,000$\,K, $32\,500$\,K and $35\,000$\,K in
temperature and $\log{g}$ of $5.0$, $5.25$ and $5.5$. The synthetic
spectra cover $3\,000$\,\AA{} to $7\,000$\,\AA{} at a resolution of
$0.1$\,\AA{}.
\begin{figure}[ht!]
\centering
\includegraphics [angle=180,scale=0.575,bb=43 14 680 267] {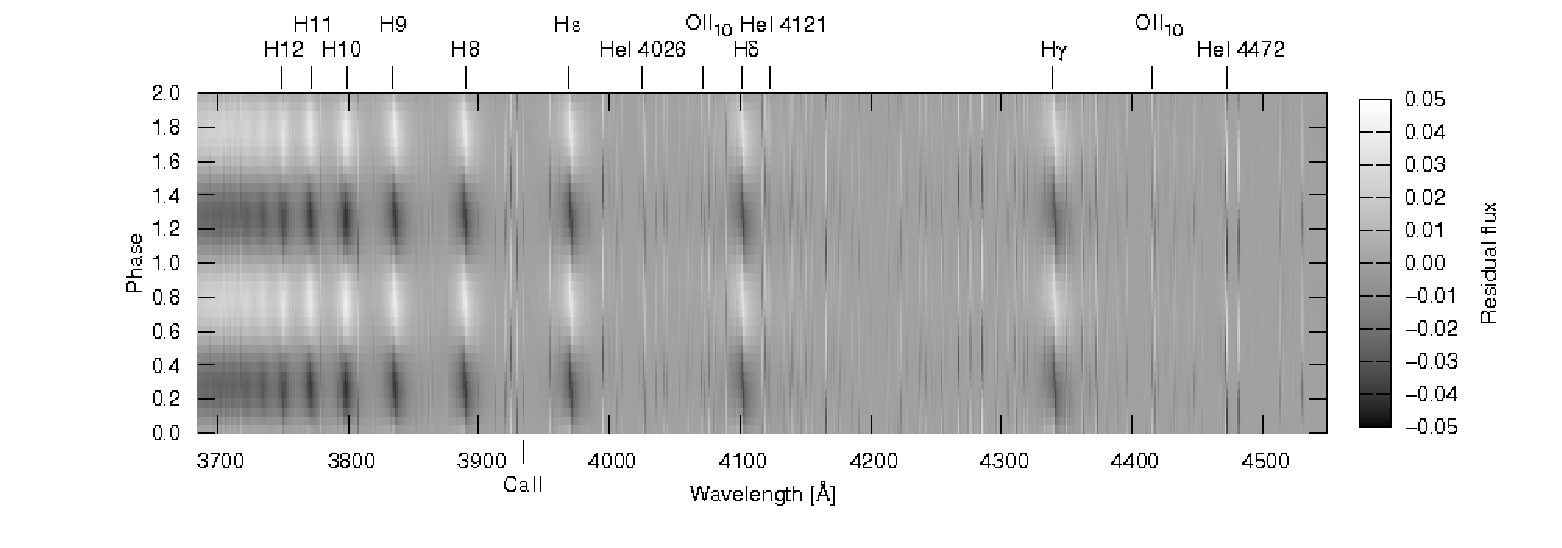}
\includegraphics [angle=270,scale=0.525]{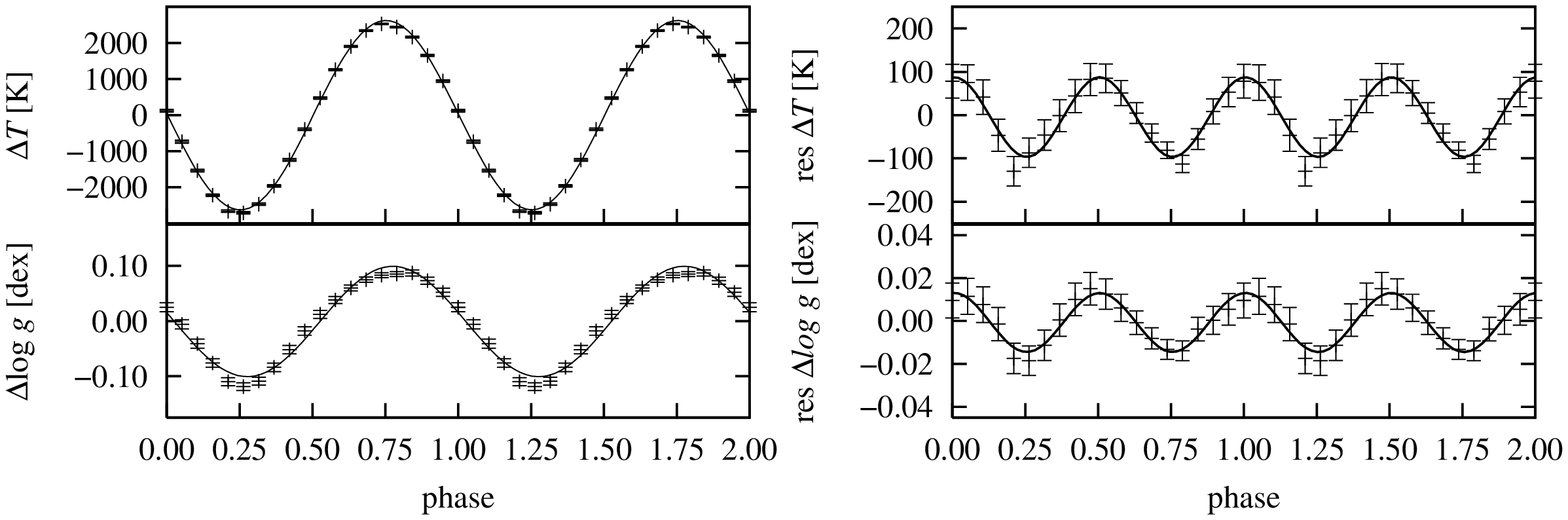}
\caption{Same as Fig.~1 but for the
synthetic line profile variations derived from our slow rotating model
{\sc i} using the same procedures as for the observed data. Note that
the first harmonic is clearly seen in both temperature and gravity at
amplitudes similar to the observed ones (compare
Fig.~\ref{fig_f1}).}\label{fig_model}
\end{figure}
Falter (2001) used a simple wavelength independent limb darkening law
in his code, which is the main drawback of this method according to
Townsend (1997). So this approximation was regarded too crude and
replaced by a quadratic law calculated for every wavelength of the
synthetic spectra. Intensity spectra for nine angels were calculated
from the model atmosphere and quadratic functions were fitted for
every wavelength.  The set of limb darkening coefficients was then
input to \textit{KYLE}.  Fig.~\ref{fig_model} (\textit{top}) shows the
theoretical line profile variation for a radial mode $l=0,m=0$ in
PG1605$+$072 calculated with the \textit{BRUCE} and \textit{KYLE}
code.  We then used the same fitting procedure described in Sect.~2 to
determine atmospheric parameters.  To complete the set of stellar
parameters we need to know the equatorial rotational velocity.  Heber
et al.~(1999) determined the projected rotational velocity to
$v_{\mathrm{rot}}\sin{i}\approx39\,\mathrm{km}\,\mathrm{s}^{-1}$ from
the observed spectral line broadening in a time integrated spectrum.
However pulsational line broadening was not considered. As the RV
amplitude of the dominant mode is large, it has to be corrected for.
Taking into account the projection factor (Montanes--Rodriguez et al.\
2001), we adopt a lower limit for the $v_{\mathrm{rot}}\sin{i}$ of
about $17.4\,\mathrm{km}\,\mathrm{s}^{-1}$.\\ Based on these
assumptions we tried two different models for PG1605+072: model {\sc
i} is a slowly rotating star with $i=90^\circ$ and
$v_{\mathrm{rot}}=17.4\,\mathrm{km}\,\mathrm{s}^{-1}$, model {\sc ii}
is a fast rotating one with $i=7.7^\circ$ and
$v_{\mathrm{rot}}=130\,\mathrm{km}\,\mathrm{s}^{-1}$. In the following
section we compare the surface variations of these very different
models.
\begin{figure}[ht!]
\centering
\includegraphics [scale=0.3] {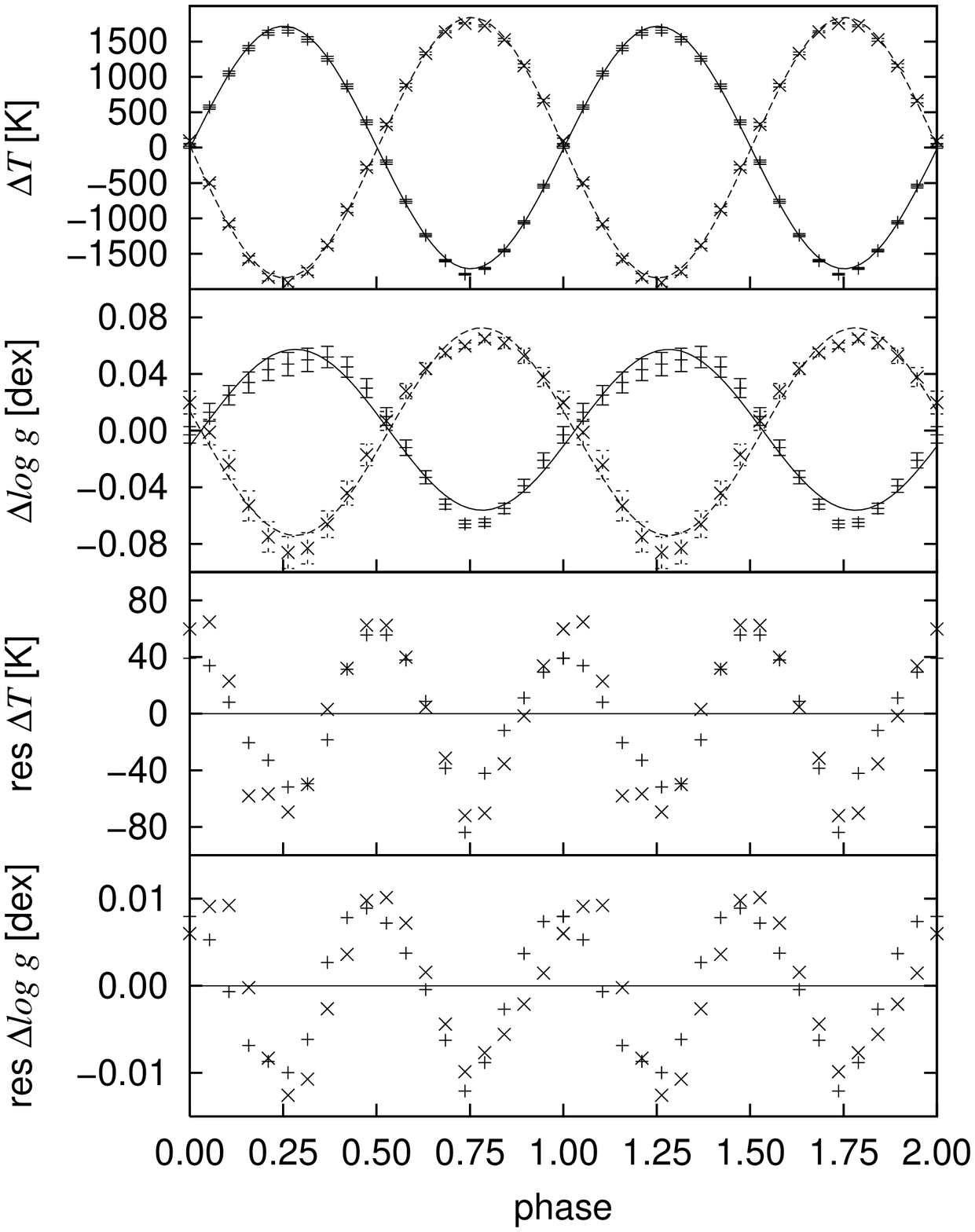}
\includegraphics [scale=0.3]{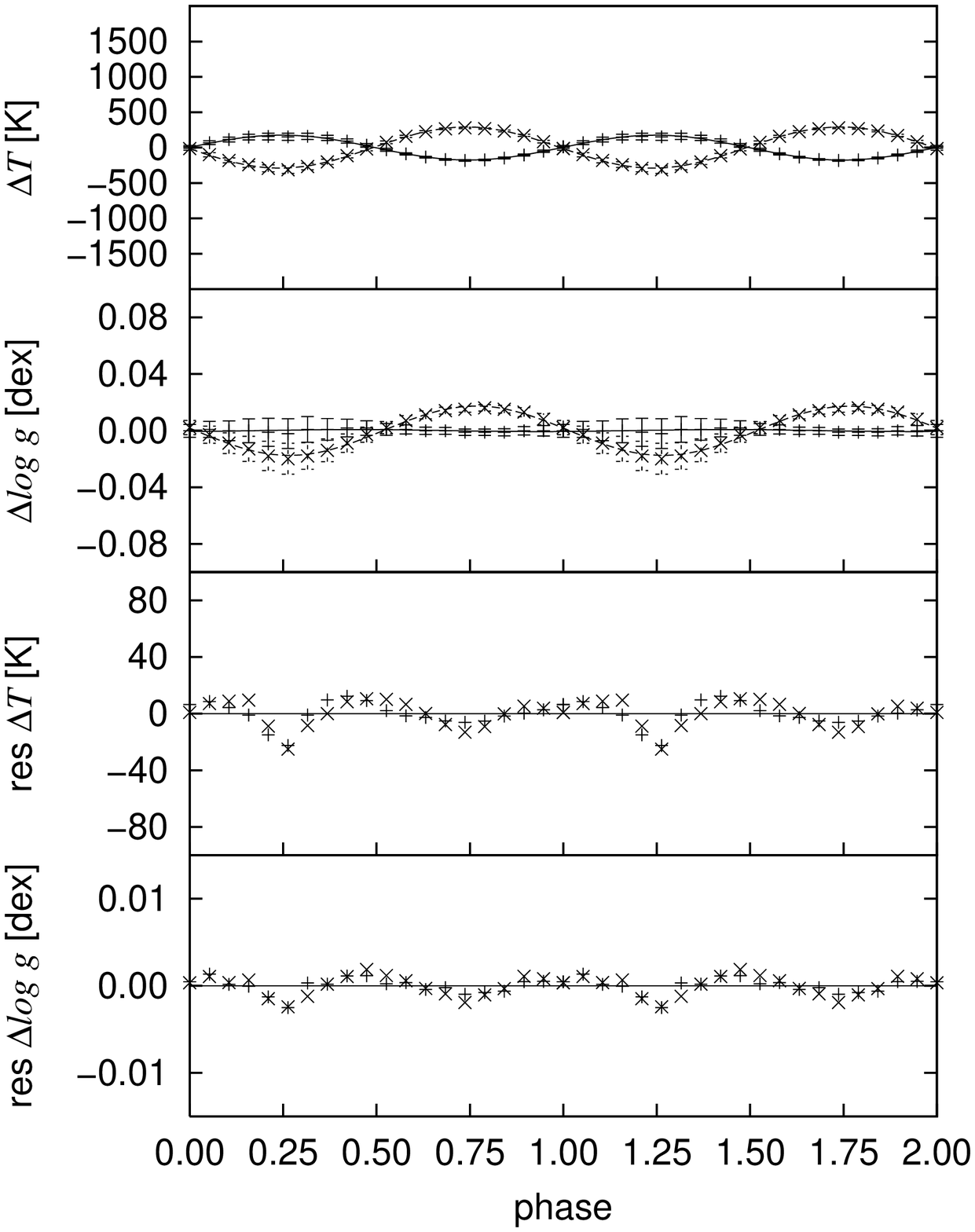}
\caption{The modelled variations of the
atmospheric parameter for the mode $l=1,m=+1$ (+) and $l=1,m=-1$ (x)
with statistical error bars, sine fits and residuals for the slowly
rotating model {\sc i} (\textit{left}) and the fast rotating model
{\sc ii} (\textit{right}).}\label{fig_l1m+-1}
\end{figure}
For the radial mode $l=0,m=0$ variations in $T_{\rm{eff}}$ and $\log{g}$ do not depend on 
inclination, due to the spherical symmetry of the problem. 
Also the residuals show variations 
of half the period attributed to the first harmonic, which have also been detected in the 
observational data at very similar amplitudes. This shows the reliability of our procedure. 
As can be seen from a comparison of Fig.~\ref{fig_f1} and Fig.~\ref{fig_model}, the predicted variations 
for $T_{\rm{eff}}$ are much larger than observed for the dominant mode, while that of the first harmonic are 
consistent with the observed ones. The predicted gravity variations however are consistent with observations 
for both the dominant and the first harmonic. In their analysis of Balloon 090100001 \O{}stensen et al.\ 
(these proceedings) showed that the temperature variations may be strongly affected by non-adiabatic effects, while 
the gravity and velocity variations are not. If we therefore rely only on the gravity variations we could conclude 
that the dominant frequency is the radial mode.

For the $l=1,m=\pm1$ modes in the slowly rotating model {\sc i} we do not see a node line any more and 
therefore the pattern is similar to a weak radial mode again. But now as we change sign on 
$m$, the sinusoidal variation shifts with a factor of $0.5$ in phase (see Fig~\ref{fig_l1m+-1}).
The amplitudes in the slow rotating model for $l=1,m=+1$ are 1\,714\,K in $T_{\rm{eff}}$ and 0.057 in $\log{g}$, 
while for $l=1,m=-1$ they are 1\,840\,K in $T_{\rm{eff}}$ and 0.073 in $\log{g}$. 
Our result is that in a slowly rotating model the prograde and retrograde pulsation modes 
look quite the same, in terms of sinusoidal variation, except for a phase shift of $0.5$.

In the fast rotating model {\sc ii} we view the star almost pole on, therefore we see the 
node line, which divides the star into 2 areas of different size. Either the hot part or the 
cold part of the star prevails and we obtain again a sinusoidal shape of the variations 
with a rather low amplitude. Our modelling shows that this is true only 
for the temperature variations. 
For $l=1,m=-1$ we measure amplitudes of 
290\,K in $T_{\rm{eff}}$ and 0.017 in $\log{g}$, while for $l=1,m=+1$ the variations are 175\,K in $T_{\rm{eff}}$ 
and 0.001 in $\log{g}$. Hence, for a retrograde mode the gravity variation almost disappears, while for the 
prograde mode it does not. 
Applying this method we derived a complete set of amplitudes for the modes $l\leq2$ (Tillich et al., in prep.).


\begin{thebibliography}{}
\bibitem{charpinet_1997}
Charpinet, S., Fontaine, G., Brassard D., et al.\ 1996, ApJ, 483, L123
\bibitem{falter_2004}
Falter, S.\ 2001, Diploma Thesis, University of Erlangen--N\"urnberg
\bibitem{heber_1999}
Heber, U., Reid, I. N., Werner K., et al.\ 1999, A\&A, 348, L25
\bibitem{jeffery_2001}
Monta{\~n}{\'e}s Rodriguez, P., \& Jeffery, C. S.\ 2001, A\&A, 375, 411
\bibitem{kilkenny_1999}
Kilkenny, D., Koen, C., O'Donoghue, D., et al.\ 1999, MNRAS, 303, 525
\bibitem{koen_1998}
Koen, C., O'Donoghue, D., Kilkenny, D., et al.\ 1998, MNRAS, 296, 317
\bibitem{napiwotzki_1999}
Napiwotzki, R. 1999, A\&A, 350, 101
\bibitem{ostensen_2007}
\O{}stensen, R., et al.\ 2007, these proceedings
\bibitem{otoole_2000}
O'Toole, S. J., Bedding, T. R., Kjeldsen, H., et al.\ 2000, ApJ, 537, L53
\bibitem{otoole_2005}
O'Toole, S. J., Heber, U., Jeffery, C. S., et al. 2005, A\&A, 440, 667
\bibitem{townsend_1997}
Townsend, R. 1997, PhD Thesis, University College London
\bibitem{tillich_2007}
Tillich, A., Heber, U., O'Toole, S. J., et al. 2007, A\&A, 473, 219
\bibitem{tillich_2007b}
Tillich, A., et al. 2008, in prep.
\end{thebibliography}
\end{document}